\documentclass[letterpaper]{article} 
\usepackage[draft]{aaai2026}  
\usepackage{times}  
\usepackage{helvet}  
\usepackage{courier}  
\usepackage[hyphens]{url}  
\usepackage{graphicx} 
\usepackage{natbib}  
\usepackage{caption} 
\usepackage{algorithm}
\usepackage{algorithmic}
\usepackage{xcolor}
\usepackage{xspace}
\usepackage{amsmath}
\usepackage{subcaption}
\newif\ifarxiv

\newif\ifcomments
\commentstrue 
\newcommand{\ourtitle}{{ConQuER}\xspace}
\newcommand{\biasname}{{data-driven}\xspace}
\usepackage{newfloat}
\usepackage{listings}
\DeclareCaptionStyle{ruled}{labelfont=normalfont,labelsep=colon,strut=off} 
\floatstyle{ruled}
\newfloat{listing}{tb}{lst}{}
\floatname{listing}{Listing}
\title{\ourtitle: Modular Architectures for Control and Bias Mitigation \\ in IQP Quantum Generative Models}
\author{
    Xiaocheng Zou\textsuperscript{\rm 1}\equalcontrib,
    Shijin Duan\textsuperscript{\rm 1}\equalcontrib,
    Charles Fleming\textsuperscript{\rm 2},
    Gaowen Liu\textsuperscript{\rm 2},
    Ramana Rao Kompella\textsuperscript{\rm 2}\\
    Shaolei Ren\textsuperscript{\rm 3},
    Xiaolin Xu\textsuperscript{\rm 1}
}
\affiliations{
    \textsuperscript{\rm 1}Northeastern University\\
    \textsuperscript{\rm 2}Cisco Research\\
    \textsuperscript{\rm 3}University of California, Riverside\\
     \{zou.xiaoc, duan.s, x.xu\}@northeastern.edu, \{chflemin, gaoliu, rkompell\}@cisco.com, sren@ece.ucr.edu
    
}

\ifarxiv
  \usepackage[hidelinks]{hyperref}
\fi

\begin{document}

\maketitle

\begin{abstract}
Quantum generative models based on instantaneous quantum polynomial (IQP) circuits show great promise in learning complex distributions while maintaining classical trainability. However, current implementations suffer from two key limitations: lack of controllability over generated outputs and severe generation bias towards certain expected patterns. We present a Controllable Quantum Generative Framework, \ourtitle, which addresses both challenges through a modular circuit architecture. \ourtitle embeds a lightweight controller circuit that can be directly combined with pre-trained IQP circuits to precisely control the output distribution without full retraining. Leveraging the advantages of IQP, our scheme enables precise control over properties such as the Hamming Weight distribution with minimal parameter and gate overhead. In addition, inspired by the controller design, we extend this modular approach through \biasname optimization to embed implicit control paths in the underlying IQP architecture, significantly reducing generation bias on structured datasets. \ourtitle retains efficient classical training properties and high scalability. We experimentally validate \ourtitle on multiple quantum state datasets, demonstrating its superior control accuracy and balanced generation performance, only with very low overhead cost $(\leq5\%)$ over original IQP circuits. Our framework bridges the gap between the advantages of quantum computing and the practical needs of controllable generation modeling.
\end{abstract}


\section{Introduction}
Quantum computing has emerged as a promising paradigm for solving computationally intractable problems, with quantum machine learning representing one of the most compelling near-term applications \cite{Preskill_2018,Biamonte2017}. Specifically, generative modeling shows particular promise, thanks to the inherent ability of quantum systems to efficiently sample from complex probability distributions \cite{doi:10.1126/sciadv.aat9004,Benedetti2019}. Recent experimental demonstrations have shown that quantum processors can achieve quantum supremacy in sampling tasks, demonstrating computational capabilities beyond classical reach \cite{Arute_2019}. Instantaneous Quantum Polynomial (IQP) circuit, a restricted model of quantum generation, has attracted significant attention in this context \cite{Bremner_2010,Bremner_2017_iqpproperty,Shepherd_bremner_2009}. The remarkable property of IQP circuits lies in their unique computational characteristics: while sampling from their output distributions is believed to be classically intractable under reasonable complexity assumptions \cite{Bremner_2010, Boixo_2018}, recent breakthroughs show that their parameters can be efficiently optimized on classical computers when combined with appropriate loss functions such as the Maximum Mean Discrepancy (MMD) \cite{fasttrainingIQP, rudolph2023trainabilitybarriersopportunitiesquantum}. Specifically, the computational effort required for parameter optimization of IQP circuits on classical computers increases only polynomially with the number of qubits and the number of gates, in contrast to the exponential scaling for general variational quantum algorithms (VQAs) where gradient computation requires simulating the full quantum state \cite{fasttrainingIQP}. This ``train on classical, deploy on quantum'' paradigm makes IQP particularly attractive for many-qubit implementation on current 
quantum devices \cite{Arute_2019}, offering a practical path toward quantum advantage in generative modeling tasks.

Despite these theoretical and practical advantages, current quantum generative models remain limited to basic distribution learning tasks. Unlike traditional VQAs that suffer from exponential training costs and barren plateaus \cite{McClean_2018, Cerezo_2021}, parameterized IQP circuits achieve remarkable scalability even over thousands of qubits. However, this scalability has not yet translated into functional sophistication. While classical generative models have evolved to support sophisticated control mechanisms, such as conditional generation in generative adversarial networks \cite{mirza2014conditionalgenerativeadversarialnets, sohn2015learning, karras2019stylebasedgeneratorarchitecturegenerative}, and guided sampling in diffusion models \cite{dhariwal2021diffusionmodelsbeatgans}, quantum models still lack comparable capabilities. Once trained, existing IQP-based models can only produce samples from the learned distribution without ability to steer or condition the generation process \cite{Dallaire_Demers_2018,Zhu_2019,fasttrainingIQP}. This limitation is particularly problematic for practical applications. For instance, in studying phase transitions using the 2D Ising model, users often need to generate spin configurations with specific properties: samples with particular magnetization values to study order parameters, configurations exhibiting specific correlation patterns to analyze critical behavior, or states representative of different phases near the transition point. Without efficient control mechanisms, quantum models must train separate models for each property from scratch, requiring substantial effort and failing to address multiple targeted generation needs simultaneously. This disparity between quantum models and classical alternatives in conditional generation severely limits the practical utility of quantum approaches.

Furthermore, original parameterized IQP circuits often exhibit severe generation bias, disproportionately favoring certain output patterns while struggling to generate others \cite{fasttrainingIQP,Du_2020}. This imbalance is particularly pronounced in structured datasets such as binary blob patterns\cite{Du_2020}, where the model may generate some patterns with high frequency, while completely failing to produce others. 
The limitations of current quantum generative models can thus be summarized as two critical shortcomings: (i) the complete absence of control mechanisms for conditional or guided generation, and (ii) severe generation bias that prevents balanced sampling across all modes of the target distribution.

To mitigate these challenges, we introduce \ourtitle (a \textbf{Con}trollable \textbf{Qu}antum G\textbf{E}nerative F\textbf{R}amework) for controllable and balanced quantum generation based on IQP circuits. \ourtitle consists of two complementary innovations that directly target the identified limitations. First, we develop a modular control mechanism that augments pre-trained generators with lightweight controller circuits. By exploiting the commutativity property of IQP gates, these controllers can steer the generation process toward desired properties, such as specific weights or correlation patterns, without changing the quantum circuit structure or requiring fully retraining. Second, 
based on the design of the controllers, 
we propose a \biasname architecture optimization strategy that analyzes the trained parameter distribution to identify key gate patterns and then implicitly embeds control structures into the original circuits. This approach significantly reduces generation bias, 
with the same computational complexity. Both solutions preserve the efficiency and scalability of classical training on IQP circuits, maintaining the potential of IQP methods to scale to quantum systems with thousands of qubits for controlled generation.

Our contributions are summarized as follows:

\begin{itemize}
    \item We propose the first systematic framework for controllable quantum generation, which adopts a modular design philosophy that enables dynamic steering of output distributions. Our approach achieves control over generated samples' properties with small overhead ($\leq5\%$).
    \item We develop a \biasname gate structure optimization method that mitigates generation bias in quantum circuits. By analyzing trained parameter magnitudes and redistributing gates based on empirical importance, our approach reduces pattern imbalance by over 18\% on structured datasets compared to the baseline IQP circuits.
    \item We use formal analysis proving that our enhancements preserve the quantum computational advantages, as well as the classical training efficiency of IQP circuits.
    \item We evaluate the performance of \ourtitle through extensive experiments on 2D Ising models and binary blob datasets, demonstrating its excellent scalability from 16 to 25 qubits with logarithmically decreasing overhead as system size increases.
\end{itemize}

\section{Background and Related Works}

\subsection{Parameterized IQP Circuits}
\begin{figure*}[htb]
    \centering
    \includegraphics[width=\textwidth]{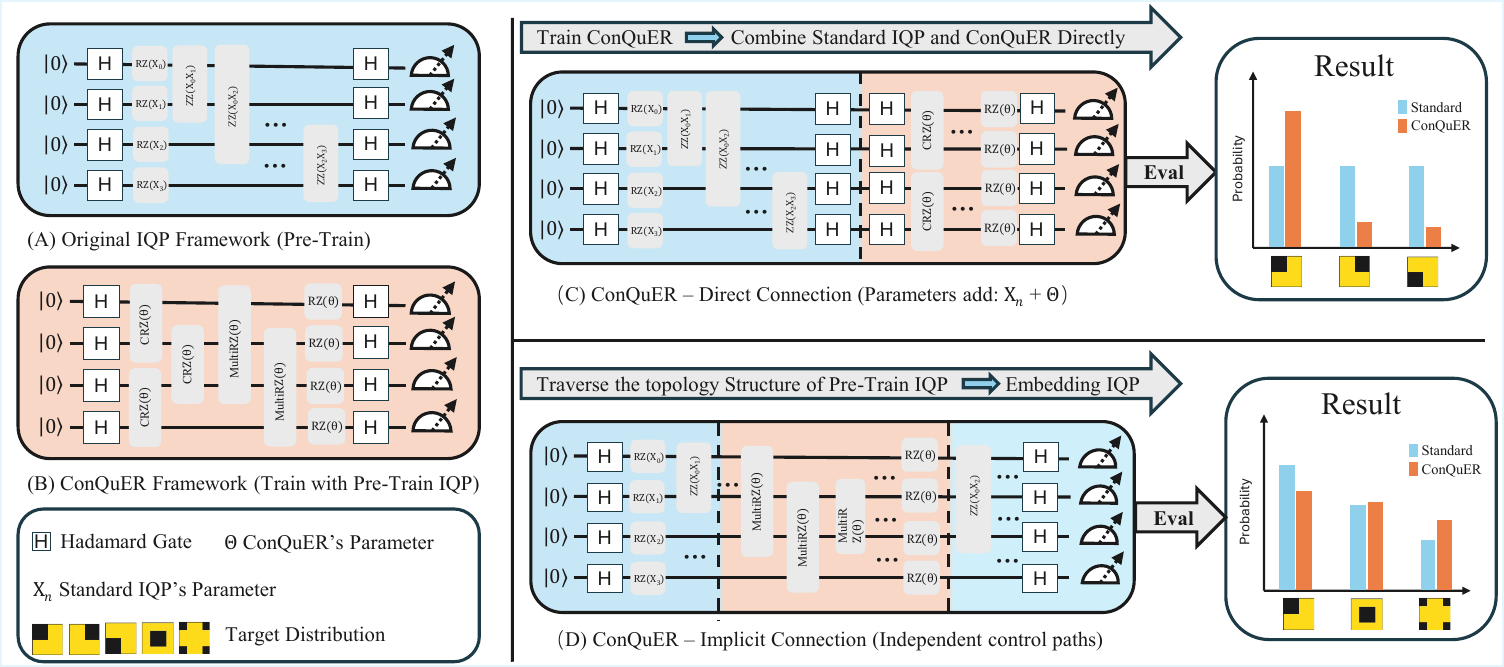}
    \caption{\ourtitle framework overview in the example of 4 qubits. (A) Original IQP circuit pre-trained on target data. (B) \ourtitle controller trained independently with the pre-trained IQP parameters. (C) Direct connection approach: controller parameters $\theta$ are added to pre-trained parameters $X_n$, enabling conditional generation. While two consecutive H gates would cancel during transpilation in quantum deployment, we retain them here for visual clarity and modularity. (D) Implicit connection approach: controller structure is embedded within the IQP topology for bias mitigation. Both approaches preserve the IQP structure while achieving different distribution control objectives.}
    \label{fig:overview}
\end{figure*}

Instantaneous Quantum Polynomial (IQP) circuits represent a restricted model of quantum computation with unique computational properties \cite{Bremner_2010,Bremner_2017_iqpproperty}. As illustrated in Figure~\ref{fig:overview}(A), an IQP circuit follows a distinctive three-layer structure: Hadamard gates on all qubits, parameterized diagonal gates in the X-basis, and final Hadamard gates before measurement. The parameterized gates take the form $\text{exp}(i\theta_j X_{g_j})$, where $X_{g_j}$ represents tensor products of Pauli-X operators acting on qubit subset $g_j$. The defining characteristic is that all these gates are diagonal in the X-basis and thus commute with each other, eliminating any temporal ordering constraints and making the computation ``instantaneous'', i.e., all gates can be applied simultaneously without regard to their order \cite{Bremner_2010, Shepherd_bremner_2009}.

IQP circuits possess a remarkable computational duality that sets them apart from general quantum circuits. While sampling from IQP output distributions is believed to be classically intractable
, the expectation values of Pauli-Z observables can be computed efficiently on classical computers \cite{VanDenNest2010}. This duality is the key to IQP's exceptional scalability. Traditional VQAs require exponential classical resources for gradient computation, limiting the optimization to circuits less than 20-30 qubits. In contrast, IQP circuits with even thousands of qubits can be trained on classical hardware ~\cite{fasttrainingIQP}.

\subsection{Quantum Generative Models and Training}
The fundamental challenge in training quantum generation lies in efficiently estimating the distance between the model and target distributions.
The Maximum Mean Discrepancy (MMD) provides an elegant solution by comparing distributions through their moments, rather than individual probabilities \cite{gretton12a}. For IQP circuits with appropriate kernel choices, the MMD loss decomposes into a weighted sum of Pauli-Z expectation values, where each term can be computed efficiently using IQP's mathematical structure. This enables gradient-based optimization without quantum hardware \cite{ rudolph2023trainabilitybarriersopportunitiesquantum}. The combination of IQP's structural properties and MMD loss has enabled model training on datasets like MNIST, which requires circuits with over 1000 qubits -- a scale unattainable with conventional quantum machine learning approaches.

However, current IQP-based generative models face two critical limitations: the absence of control mechanisms for conditional generation, and severe generation bias on structured datasets. These limitations motivate our development of the \ourtitle framework.

\section{Our Proposed Method: \ourtitle}
This section presents \ourtitle, a unified framework for controllable quantum generation through distribution controllers. Our method  addresses the fundamental limitations in current quantum generative models by developing modular controllers that can perform both conditional generation and bias mitigation on IQP circuits. Rather than proposing separate solutions for each challenge, we introduce a unique architectural principle. It leverages the mathematical properties of IQP circuits to enable precise control over quantum distributions while simultaneously maintaining the training efficiency of IQP on classical computers.

\subsection{\ourtitle: Design Motivation and Principles} 
Traditional quantum generative models, once trained, produce determined output distributions that cannot be adjusted unless after retraining. This limitation severely restricts their practical utility, as real-world applications often require generating samples with specific properties or ensuring balanced coverage across different conditions.

Our solution introduces the concept of distribution controller, i.e., a lightweight quantum circuit in conjunction with pre-trained IQP generators to modulate their output behavior. The controller exploits a crucial property of IQP circuits: \textbf{\textit{All gates are diagonal in the X-basis and, therefore, commute}}. This commutativity enables us to add control mechanisms without destroying the quantum structure or requiring architectural changes to the base generator. The controller acts as a quantum ``router'' that can focus, redirect, or rebalance the output distribution according to specific objectives. Figure~\ref{fig:overview} illustrates the integration of our controllers with base IQP circuits through parameter additivity (direct connection) or structural embedding (implicit connection).

The design philosophy resides on three key principles. \textbf{Flexibility:} modularity ensures that controllers can be developed independently and combined with different generators. \textbf{Parameter Efficiency:} the control mechanism only adds minimal computational overhead. \textbf{Scalability:} preservation of classical trainability maintains the scalability advantages of IQP circuits.
These principles guide our architectural choices and ensure that both conditional generation and bias mitigation can be achieved within the same framework. Below, we prove the principles of efficiency and scalability of our proposed controller.

\subsection{\ourtitle: Theoretical  Foundation}
\textbf{Maintaining Classical Training Efficiency.}
The feasibility of our controller framework critically depends on maintaining the efficient classical training properties of IQP circuits \cite{fasttrainingIQP}. Original IQP circuits admit polynomial-time computation of expectation values.
We prove that our controller preserves this property.
Consider an IQP circuit implementing the unitary:
\begin{equation}
    U_\text{IQP}(\theta)=H^{\otimes n}\left(\prod_{j=1}^{m} \exp \left(i \theta_j X_{g_j}\right)\right) H^{\otimes n}
\end{equation}
where $X_{g_j}$ denotes a tensor product of Pauli-X operators on qubit subset $g_j$, and $j$ indexes the $m$ gates in the circuit. The defining property is that all parameterized gates commute: $[X_{g_i}, X_{g_j}] = X_{g_i}X_{g_j} - X_{g_j}X_{g_i} = 0$ for all $i, j$. This commutativity is preserved when we introduce our controller circuit $U_\text{ctrl}(\phi)$ with the same IQP structure.

As a result, the integrated system implements:
\begin{equation}
    U_\text{combined} (\theta, \phi) = U_\text{ctrl}(\phi) \cdot U_\text{IQP}(\theta)
\end{equation}
Due to commutativity, gates acting on identical qubit subsets combine through parameter addition in the exponent:
\begin{equation}
    \text{exp}(i\phi_{k}X_{g_k}) \cdot \text{exp}(i\theta_kX_{g_k}) = \text{exp}(i(\phi_k + \theta_k)X_{g_k})
\label{eq:param_addition}
\end{equation}
where $k$ indexes gates that act on the same qubit subset $g_k$ in both the controller and base IQP circuits. When controller gate $k'$ and base gate $k$ share the same subset (i.e., $g_{k'} = g_k$), their parameters combine additively, where the key insight enabling our control mechanism. The controller parameters $\phi$ directly modulate the effective parameters of the combined circuit without changing its fundamental structure. Crucially, the combined circuit remains an IQP circuit, preserving all computational advantages.

\textbf{MMD Loss Decomposition and Scalability.}
Training quantum generative models typically involves minimizing the MMD between the model and target distributions. For IQP, the MMD loss admits an efficient decomposition:

\begin{equation}
    \text{MMD}=\sum_k \alpha_k\left\langle Z_{S_k}\right\rangle_ \text{combined }
\end{equation}
where $\langle Z_S \rangle$ denotes the expectation value of Pauli-Z operators acting on qubit subset $S$, and the sum runs over polynomially many subsets. The coefficient $\alpha_k$ depends on the kernel choice and training data. This decomposition is efficient because: (1) only polynomially many terms contribute non-zero coefficients, and (2) each expectation value $\langle Z_S \rangle$ can be computed in polynomial time using graph-theoretic methods specific to IQP circuits \cite{VanDenNest2010}.

For our combined controller-generator system, we prove that this efficient decomposition is preserved. Let $\theta_{eff} = \theta + \phi$ be the effective parameters, following Eq.~\ref{eq:param_addition}. The expectation value computation remains:
\begin{equation}
\begin{aligned}
& \left\langle Z_S\right\rangle_{\text {combined }}= \\
& \left\langle 0^n\right| H^{\otimes n}\left(\prod_j \exp \left(-i \theta_{e f f, j} X_{g_j}\right)\right) H^{\otimes n} \\
& Z_S H^{\otimes n}\left(\prod_j \exp \left(i \theta_{e f f, j} X_{g_j}\right)\right) H^{\otimes n}\left|0^n\right\rangle
\end{aligned}
\end{equation}
This expression maintains polynomial-time computability, enabling training on hundreds of qubits where traditional VQAs fail due to exponential resource requirements. The scalability stems from IQP's special structure that allows classical computation of quantum expectation values without simulating the full quantum state. Specifically, while VQAs require $O(2^n)$ operations to compute gradients, IQP circuits with MMD loss require only $O(\text{poly}(n))$ operations, where the polynomial degree depends on the maximum gate order (typically 6 for practical implementations). 

\textbf{Controlled-RZ Transformation for Enhanced Control.}
While IQP gates are traditionally expressed as $\exp(i\theta X_{g})$ rotations, we leverage an equivalent representation using controlled-RZ (CRZ) gates to enhance control precision. For two-qubit gates, the transformation is:
\begin{equation}
    \text{exp}(i\theta X_1X_2) = CNOT_{1,2} \cdot R_Z^2(2\theta) \cdot CNOT_{1,2}
\end{equation}
This CRZ representation provides significant advantages for our controller design. Straightforwardly, it makes explicit how two-qubit IQP gates introduce phase relationships conditioned on computational basis states. Therefore, it enables more intuitive reasoning about how parameter changes affect the quantum state. Furthermore, it facilitates the design of controllers that target specific correlation patterns in the output distribution.

The transformation from MultiRZ to CRZ gates extends the multi-qubit gates through decomposition into two-body interactions, maintaining the efficient classical simulation properties while providing more fine-grained control granularity. As shown in Figure \ref{fig:overview}(B), this representation guides our controller architecture development.

\textbf{Parameter Influence on Output Distributions.} 
Understanding how individual gate parameters influence the output distribution is crucial for effective controller design. We analyze this relationship through first-order perturbation theory. For a small parameter change $\delta\theta_k$ in gate $k$, the change in output probability for bitstring $x$ is:
\begin{equation}
    \delta p(x) \approx 2 \operatorname{Re}\left[\langle x| U^{\dagger} \frac{\partial U}{\partial \theta_k}\left|0^n\right\rangle\left\langle 0^n\right| U^{\dagger}|x\rangle\right] \delta \theta_k
\end{equation}
this reveals a hierarchical influence structure: single-qubit gates primarily control marginal bit probabilities $p(x_i = 1)$, two-qubit gates manage pairwise correlations $p(x_i=a, x_j=b)$, and higher-order gates capture complex dependencies with diminishing direct influence.

\subsection{Controller Architecture and Implementation}
Based on our mathematical analysis and empirical parameter studies, we present the controller architecture that implements our distribution control mechanism. To further validate our theoretical predictions about parameter influence, we first analyze the learned parameters of pre-trained IQP circuits on a 16-qubit system as a case study.

\begin{figure}[t]
  \centering
  \begin{subfigure}{0.23\textwidth}
    \centering
    \includegraphics[width=\linewidth]{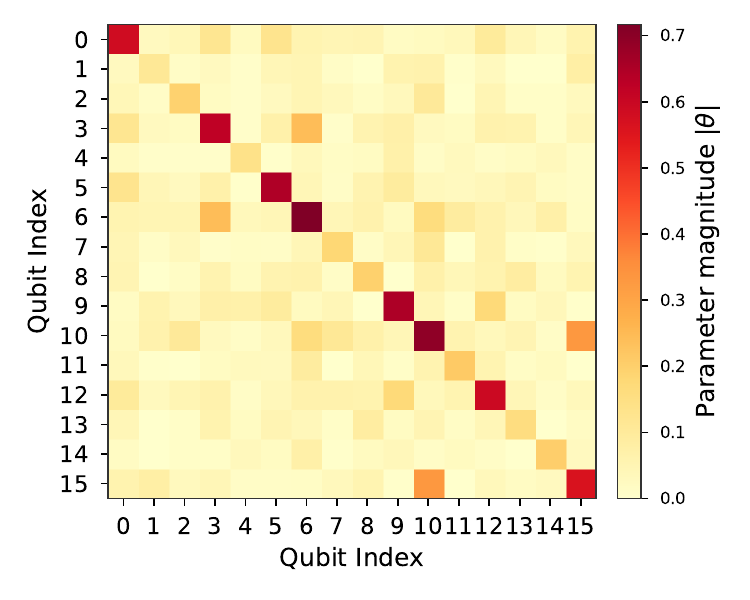}
    \caption{}
    \label{fig:heatmap1}
  \end{subfigure}
  \hfill
  \begin{subfigure}{0.23\textwidth}
    \centering
    \includegraphics[width=\linewidth]{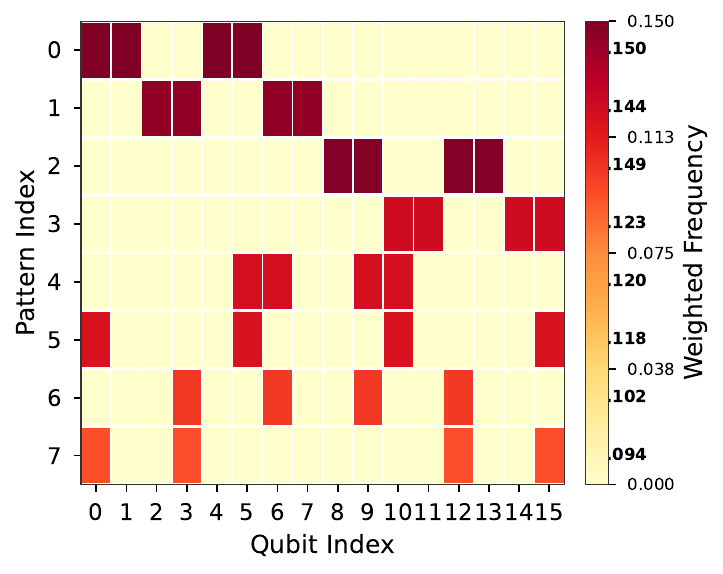}
    \caption{}
    \label{fig:heatmap2}
  \end{subfigure}

  \caption{Parameter influence analysis on 16-qubit systems. (a) Learned parameter magnitudes across qubit pairs in pre-trained IQP circuit. (b) Qubit-wise parameter importance for each of 8 binary patterns. Darker colors indicate larger values.}
  \label{fig:heatmap}
\end{figure}

Figure~\ref{fig:heatmap}(a) reveals the learned parameter importance across different qubit pair connections, with diagonal patterns showing the strongest values. Our analysis confirms that single-qubit gates exhibit larger average parameter magnitudes (0.40) compared to two-qubit gates (0.04), validating our theoretical analysis. Figure~\ref{fig:heatmap}(b) demonstrates how these parameter patterns directly correlate with generation bias—patterns with higher cumulative parameter weights are generated more frequently, providing empirical evidence for the bias we aim to mitigate.

\begin{algorithm}[tb]
\caption{ConQuER Controller Construction}
\label{alg:conquer-construction}
\begin{algorithmic}[1]
  \STATE \textbf{Input:} number of qubits $n$, control objective $\mathsf{obj}$, base IQP params $\theta_{\text{base}}$
  \STATE \textbf{Output:} controller circuit $\mathcal{C}_{\text{ctrl}}$, controller parameters $\phi$
  \STATE $G_{\text{ctrl}} \leftarrow [\,]$
  \FOR{layer $\in \{\text{even},\text{odd}\}$}\label{alg:line4}
      \STATE $\mathcal{I} \leftarrow \begin{cases}
          \{0, 2, 4, \ldots, n-2\} & \text{if layer = even}\\
          \{1, 3, 5, \ldots, n-3\} & \text{if layer = odd}
      \end{cases}$
      \FOR{$i \in \mathcal{I}$}
          \STATE $G_{\text{ctrl}} \mathrel{+}= \operatorname{TwoQubitGate}(i,i+1)$
      \ENDFOR
  \ENDFOR\label{alg:line8}
  \STATE $G_{\text{ctrl}} \leftarrow G_{\text{ctrl}} \cup \textsc{GetControlGates}(\mathsf{obj}, n, \theta_{\text{base}})$\label{alg:line10}
  \FOR{$i = 0$ \textbf{to} $n-1$} \label{alg:line11}
      \STATE $G_{\text{ctrl}} \mathrel{+}= \operatorname{SingleQubitGate}(i)$
  \ENDFOR\label{alg:line13}
  \STATE $\phi \leftarrow \textsc{SmartInitialize}(G_{\text{ctrl}}, \mathsf{obj})$
  \STATE \textbf{return} $(\operatorname{IQPCircuit}(G_{\text{ctrl}}, n), \phi)$
\end{algorithmic}
\end{algorithm}

\begin{algorithm}[tb]
\caption{Objective-Specific Control Gate Selection}
\label{alg:control-gates}
\begin{algorithmic}[1]
  \STATE \textbf{Function} \textsc{GetControlGates}($\mathsf{obj}$, $n$, $\theta_{\text{base}}$)
  \STATE $G_{\text{extra}} \leftarrow [\,]$
  \IF{$\mathsf{obj} \in \{\text{``high\_weight''}, \text{``low\_weight''}\}$}
      \FOR{$i = 0$ \textbf{to} $n-3$}
          \STATE $G_{\text{extra}} \mathrel{+}= \operatorname{TwoQubitGate}(i,i+2)$
      \ENDFOR
  \ELSIF{$\mathsf{obj} = \text{``bias\_mitigation''}$} \label{alg:2line7}
      \STATE $\mathcal{W} \leftarrow \textsc{ExtractWeights}(\theta_{\text{base}})$
      \FOR{$(i,j)$ where $\mathcal{W}_{i,j} < 0.1$ and $|i-j| \leq 3$}\label{alg:2line9}
          \STATE $G_{\text{extra}} \mathrel{+}= \operatorname{TwoQubitGate}(i,j)$
      \ENDFOR\label{alg:2line11}
      \FOR{$i$ where $\sum_j \mathcal{W}_{i,j} < \text{avg}(\mathcal{W})$}
          \STATE $G_{\text{extra}} \mathrel{+}= \operatorname{SingleQubitGate}(i)$
      \ENDFOR\label{alg:2line14}
  \ENDIF
  \STATE \textbf{return} $G_{\text{extra}}$
\end{algorithmic}
\end{algorithm}

Algorithm~\ref{alg:conquer-construction} presents our controller construction with a three-layer architecture. Layer 1 (lines \ref{alg:line4}-\ref{alg:line8}) implements an alternating pattern of nearest-neighbor two-qubit gates, first connecting even-leading qubit pairs $(0,1), (2,3), \ldots$, then odd-leading pairs $(1,2), (3,4), \ldots$. This pattern ensures full connectivity between adjacent qubits while maintaining a constant circuit depth of 2. Layer 2 (line \ref{alg:line10}) adds objective-specific gates based on the control task through \textsc{GetControlGates}, detailed in Algorithm~\ref{alg:control-gates}. Layer 3 (lines \ref{alg:line11}-\ref{alg:line13}) places single-qubit gates on all qubits, providing direct control over bit-wise statistics. 

The parameter initialization strategy (\textsc{SmartInitialize} in Algorithm~\ref{alg:conquer-construction}) directly follows our theoretical analysis. For high-weight control, single-qubit parameters initialize with negative values around -0.1 to promote bit flips from $|0\rangle$ to $|1\rangle$. For low-weight control, positive values around 0.1 suppress these transitions. Balanced generation uses small Gaussian noise $\mathcal{N}(0, 0.01)$ around zero. This principled initialization accelerates convergence by starting the optimization closer to the desired solution.

\textbf{Conditional Generation.}
In conditional generation, we leverage the modular controller design to steer the pre-trained generator toward specific output distributions. The key insight from parameter additivity analysis (Eq.~\ref{eq:param_addition}) is that the controller only needs to learn the residual transformation between the base distribution and the target conditional.

For a target property, e.g., high Hamming weight, we construct the controller following Algorithm~\ref{alg:conquer-construction} with $\mathsf{obj} = \text{``high\_weight''}$. This adds next-nearest neighbor connections (Algorithm~\ref{alg:control-gates}, lines \ref{alg:2line9}-\ref{alg:2line11}) to enhance correlations between distant qubits, promoting the generation of bitstrings with many 1s. The training process creates a filtered dataset $\mathcal{D}_{\text{target}}$ containing only samples satisfying the target condition, then optimizes the controller parameters to minimize:
\begin{equation}
    \mathcal{L}_\text{cond} = \text{MMD}(p_\text{combined}(\theta + \phi), p_\text{target})
\end{equation}

In terms of efficiency, the sparse controller only adds minimal overhead. For the example of a 16-qubit system, it adds only 45 parameters (16 single-qubit + 15 nearest-neighbor + 14 next-nearest-neighbor) compared to base IQP circuit's thousands of parameters.

\textbf{Bias Mitigation.}
Figure~\ref{fig:heatmap}(b) reveals that base IQP circuits favor certain patterns while underrepresenting others. 
Our data-driven bias mitigation leverages the parameter magnitude analysis to identify and correct these imbalances.
Algorithm~\ref{alg:control-gates} (lines \ref{alg:2line7}-\ref{alg:2line14}) implements the bias mitigation strategy by analyzing the pre-trained parameter weights $\mathcal{W}$ extracted from the heatmap analysis. The controller adds gates to underrepresented connections (those with weights below 0.1) and places additional single-qubit gates on qubits with below-average total connection strength. This implicit architectural redesign redistributes the importance across the circuit without changing the base generator.

The training objective incorporates a variance penalty to encourage balanced generation:
\begin{equation}
    \mathcal{L}_\text{bias} = \text{MMD}(p_\text{combined}, p_\text{data}) + \lambda \cdot \text{Var}(p_\text{modes})
\end{equation}
where $\text{Var}(p_\text{modes})$ measures the variance in generation frequency across different output patterns. The controller learns to counteract the inherent biases, resulting in more uniform coverage of distributions.

Please note that both scenarios leverage the same mathematical framework of parameter additivity and efficient optimization, differing only in their training objectives and gate selection strategies.

\begin{figure}
  \centering
  \begin{subfigure}[t]{0.225\textwidth}
      \centering
      \includegraphics[width=\linewidth]{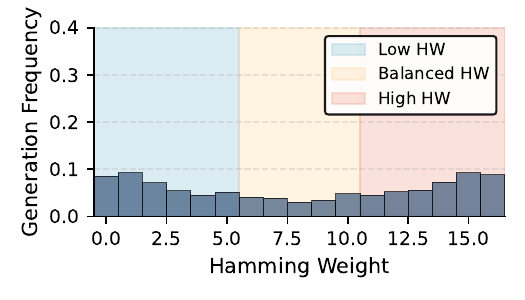}
      \caption{Baseline IQP Distribution}
      \label{fig:ising_result1}
  \end{subfigure}
  \hfill
  \begin{subfigure}[t]{0.225\textwidth}
      \centering
      \includegraphics[width=\linewidth]{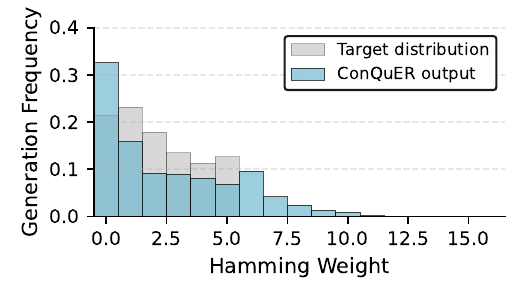}
      \caption{ConQuER - Low HW}
      \label{fig:ising_result2}
  \end{subfigure}

  \begin{subfigure}[t]{0.225\textwidth}
      \centering
      \includegraphics[width=\linewidth]{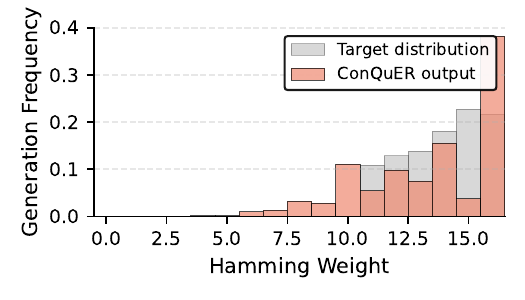}
      \caption{ConQuER - High HW}
      \label{fig:ising_result3}
  \end{subfigure}
  \hfill
  \begin{subfigure}[t]{0.225\textwidth}
      \centering
      \includegraphics[width=\linewidth]{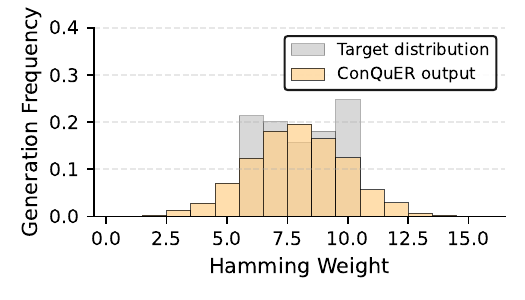}
      \caption{ConQuER - Balance HW}
      \label{fig:ising_result4}
  \end{subfigure}

  \caption{Hamming weight (HW) control on 2D Ising model. (a) Baseline IQP shows bimodal distribution. (b-d) ConQuER controls output distributions: (b) Low HW mode (81.36\% in 0-5 range), (c) High HW mode (80.5\% in 11-16 range), (d) Balanced mode (centered at weight 8, $\sigma$=1.99).}
  \label{fig:ising_result}
\end{figure}

\section{Experiment}

\subsection{Setup}
\textbf{Benchmark.} Our experimental evaluation focuses on comparing \ourtitle to a standard parameterized IQP baseline, which represents the SOTA in scalable quantum generative models. This comparison strategy is both deliberate and necessary. While numerous quantum generative models exist to date, such as variational quantum circuits and quantum GANs, they rely on gradient-based optimization\cite{quantumnat} or parameter-shift updating \cite{parametershift}, requiring exponentially large classical resources. 
In contrast, IQP circuits trained with MMD can efficiently and stably process more qubits in classical computing, making them the only viable baseline for evaluating large-scale quantum generative tasks. Furthermore, to the best of our knowledge, \textbf{no prior work has addressed the specific challenges of controllability and bias mitigation in quantum generative models}. Therefore, our experiments necessarily focus on demonstrating improvements over the original IQP (i.e., baseline in Figure~\ref{fig:ising_result}), as it represents both the most scalable approach and a direct predecessor to our contributions.
\textbf{Quantum Generative Tasks.} We evaluate \ourtitle on two quantum generative tasks that represent different challenges in quantum machine learning. The first task involves generating samples from the 2D Ising model, a fundamental model in statistical physics that exhibits rich phase transition behavior. The second task uses a binary blob dataset consisting of eight distinct 16-bit patterns with structured spatial correlations. Please refer to the Appendix for further detail.
\textbf{Environment.} All experiments are conducted on an Intel Xeon 6442Y CPU (2.6GHz, 512GB RAM). Quantum circuit simulations are run using PennyLane-0.41.1\cite{pennylane}.

\subsection{Controllable Generation}
Figure~\ref{fig:ising_result1} demonstrates \ourtitle's precise control over the Hamming weight distribution across three target modes on the 16-bit 2D Ising model. The baseline IQP circuit (Figure~\ref{fig:ising_result1}) exhibits a characteristic bimodal distribution with peaks at both extremes (weights 0-2 and 14-16), reflecting the system's natural tendency to favor fully aligned states—either all spins down (anti-ferromagnetic) or all spins up (ferromagnetic). This bimodal behavior aligns with the physics of the Ising model at low temperatures, where the system prefers ordered states over mixed configurations.

\ourtitle successfully reshapes this distribution according to control objectives. For low Hamming weight control (Figure~\ref{fig:ising_result2}), the controller concentrates 81.36\% of generated samples in the 0-5 weight range, effectively steering generation toward anti-ferromagnetic configurations. 
Conversely, for high Hamming weight control (Figure~\ref{fig:ising_result3}), 80.5\% of samples fall within the 11-16 range, demonstrating equally effective control toward ferromagnetic states through negative parameter initialization that promotes bit flips.

The balanced mode (Figure~\ref{fig:ising_result4}) presents the most challenging control task, requiring the controller to counteract the system's natural bimodal tendency and generate samples centered around intermediate weights. \ourtitle successfully creates a unimodal distribution centered at weight 8 with standard deviation $\sigma=1.99$, effectively populating the critical region where the baseline model struggles to access. This demonstrates the controller's ability to not only shift distributions but fundamentally reshape them, enabling exploration of physically interesting regimes that standard IQP circuits underrepresent.

The precision of control across all three modes validates our theoretical framework. The parameter additivity property (Eq.~\ref{eq:param_addition}) allows the lightweight controller to modulate the pre-trained generator's behavior without destroying the learned distribution structure, confirming that our modular approach can efficiently implement diverse control objectives without retraining the base model.

\begin{figure}
  \centering
  \begin{subfigure}{0.5\textwidth}
    \centering
    \includegraphics[width=\linewidth]{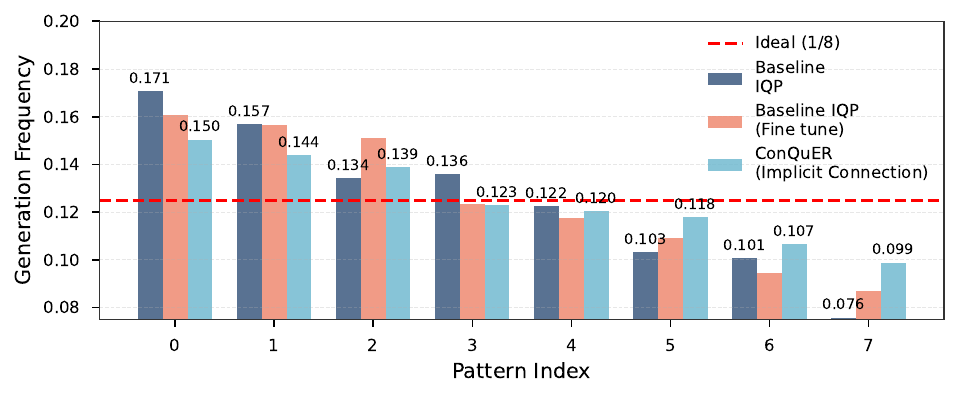}
    \caption{Pattern Generation Frequency}
    \label{fig:sub-a}
  \end{subfigure}
  \hfill
  \begin{subfigure}{0.5\textwidth}
    \centering
    \includegraphics[width=\linewidth]{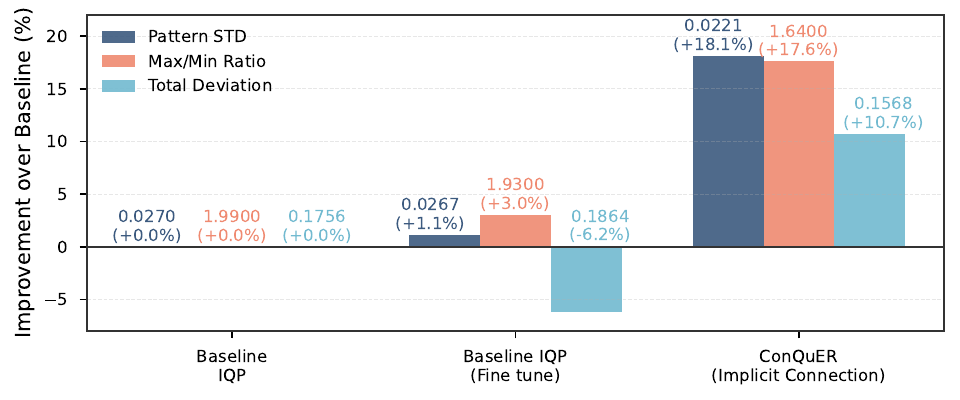}
    \caption{Quantitative Improvement Metrics}
    \label{fig:sub-b}
  \end{subfigure}

  \caption{Bias mitigation on binary blob dataset. ConQuER (implicit connection) achieves more uniform distribution across 8 patterns compared to baseline IQP. ConQuER reduces pattern STD by 18.1\%, max/min ratio by 17.6\%, and total deviation by 10.7\%.}
  \label{fig:blob_result}
\end{figure}

\subsection{Generation Bias Mitigation}
The baseline IQP circuit exhibits severe generation bias on the binary blob dataset, as shown in Figure~\ref{fig:sub-a}. Pattern 0 appears with frequency 0.171 while pattern 7 occurs with only 0.076 frequency, a 2.25× imbalance between maximum and minimum frequencies. 

Recalling our parameter analysis in Figure~\ref{fig:heatmap}, which reveals the source of such bias: single-qubit gates with significantly larger parameter magnitudes (up to 0.40) compared to two-qubit gates (below 0.15). The correlation between these parameter patterns and generation frequency (Figure~\ref{fig:heatmap2}) directly links the architectural bias to the output distribution imbalance, yet why patterns with higher cumulative parameter weights are generated more frequently.

Using the \biasname optimization strategy described for bias mitigation, \ourtitle identifies and enhances underrepresented connections through the controller architecture. The results in Figure~\ref{fig:sub-b} demonstrate substantial improvements across all bias metrics. The pattern standard deviation reduces by 18.1\% (from 0.0270 to 0.0221), indicating significantly more balanced generation across all eight patterns. The max/min ratio improves by 17.6\% (from 1.99 to 1.64), directly addressing the severe disparity between most and least frequent patterns. The total deviation from ideal uniform distribution decreases by 10.7\%.

These improvements validate our hypothesis that generation bias in quantum circuits stems from architectural choices rather than fundamental limitations. The implicit connection approach successfully redistributes importance across the circuit, achieving more uniform pattern generation without adding parameters or changing the base model. This demonstrates that empirically-guided structural modifications can effectively address longstanding bias issues in quantum generative models.

\begin{figure}
    \centering
    \includegraphics[width=\linewidth]{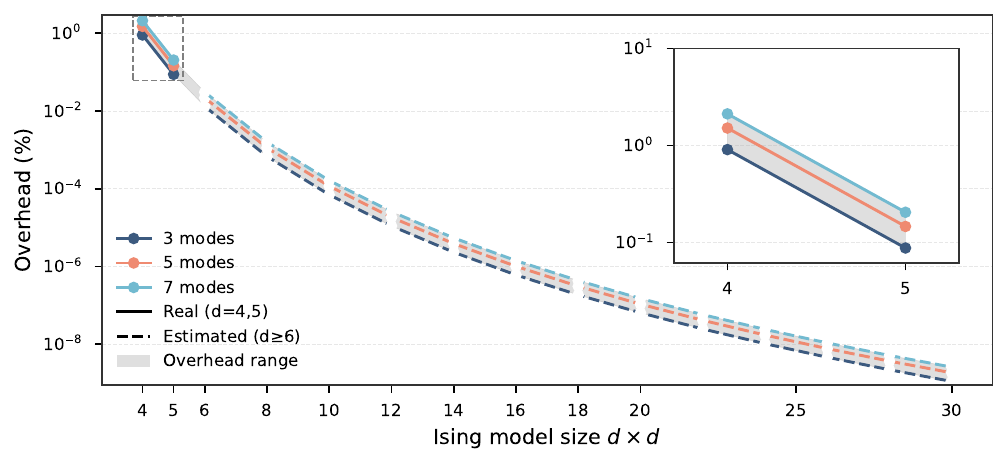}
    \caption{Controller overhead v.s. Ising model size for different numbers of control modes. Solid lines show measured data for 4×4 and 5×5 models, dashed lines show projections for larger systems. The main plot demonstrates logarithmic decrease in overhead percentage as system size increases. Inset shows the same data on a linear scale for the measured systems. }
    \label{fig:overhead}
\end{figure}
\subsection{Controller Efficiency}
For a 16-qubit system (4×4 Ising model), the base IQP circuit contains 14,892 parameters, while each controller adds only 45 parameters, resulting in a mere 0.3\% overhead. In practice, deploying three control modes (e.g., low, high, and balanced Hamming weight) requires an additional 135 parameters (overhead 0.9\%), deploying five modes requires an additional 225 parameters (overhead 1.5\%), and deploying seven modes requires an additional 315 parameters (overhead 2.1\%). Therefore, our proposed low-overhead approach enables the \textbf{``train once, control many''} paradigm: a pre-trained base model can support multiple control objectives without retraining. For a circuit with max weight of 6, the controller converges in approximately 500 iterations, while the base model requires over 2,000, achieving a $4\times$ training speedup.

Figure ~\ref{fig:overhead} demonstrates the excellent scalability of \ourtitle at various system sizes. Using real data from 4×4 and 5×5 Ising models, we observe that the percentage overhead decreases logarithmically with system size. For the tested 25-qubit system (5×5 model), even with seven control modes, the overhead increases by less than 1.5\%. Predictions for larger systems show that this trend persists -- even with 7 modes, the overhead for a 30×30 system drops to less than 0.1\%. This high scalability is due to the linear growth of controller parameters ($O(n)$) while the underlying IQP parameters grow polynomially ($O(n^6)$), when the maximum weight is 6. The shaded area indicates the range of overhead for different numbers of control modes, confirming that \ourtitle maintains excellent efficiency at all scales.

\section{Conclusion and Future Work}
This work proposes \ourtitle, a modular framework for controllable quantum generation that addresses critical limitations in current quantum generative models. By exploiting IQP's mathematical properties, our lightweight controllers achieve precise conditional generation and 18\% bias reduction with less than 5\% parameter overhead. \ourtitle inherits IQP's benefits like polynomial-time classical training and scaling beyond traditional VQAs' limitations, demonstrating that superior controllability and balanced generation are achievable in near-term quantum systems. Looking forward, developing more fine-grained control mechanisms could enable precise manipulation of specific quantum correlations beyond Hamming weight, such as entanglement patterns, local magnetization distributions, or higher-order quantum correlations. Additionally, enhancing the expressive power of IQP circuits through hybrid architectures that strategically incorporate non-commuting gates at specific circuit positions could capture more complex quantum phenomena while preserving the scalability essential for practical quantum machine learning applications.

\bibliography{ref/aaai2026}
\end{document}